\documentclass[reqno]{amsart}%
\usepackage{amsfonts}
\usepackage{amsmath}
\usepackage{amssymb}
\usepackage{graphicx}%
\theoremstyle{plain}
\newtheorem{theorem}{Theorem}[section]

\newtheorem{definition}{Definition}

\newtheorem{lemma}[theorem]{Lemma}

\newtheorem{proposition}{Proposition}

\usepackage{setspace}
\numberwithin{equation}{section}
\begin{document}
\title[ ]{Integrals of Irreducible Representations of Classical Groups}
\author[]{Da Xu and  Palle Jorgensen}
\address{Department of Mathematics\\
The University of Iowa\\
Iowa City, IA 52242} \email{dxu@math.uiowa.edu,jorgen@math.uiowa.edu
}
\maketitle

\begin{abstract}This paper is concerned with integrals which integrands are the monomials of matrix elements of irreducible representations of classical groups. Based on analysis on Young tableaux, we discuss some related duality theorems and compute the asymptotics of the group integrals when the signatures of the irreducible representations are fixed, as the rank of the classical groups go to infinity. These group integrals have physical origins in quantum mechanics, quantum information theory, and lattice Gauge theory.
\end{abstract}

% ----------------------------------------------------------------
\section{Some origins in physics}
In the present paper, we shall consider the group integral
\begin{align}
\int_G  \rho^{\lambda^{(1)}}_{i_1 j_1}\rho^{\lambda^{(2)}}_{ i_2 j_2}\cdots \rho^{\lambda^{(k)}}_{i_k j_k} \bar{\rho}^{\lambda^{'(1)}}_{ i'_1  j'_1}\bar{\rho}^{\lambda^{'(2)}}_{ i'_2 j'_2}\cdots \bar{\rho}^{\lambda^{'(k^{'})}}_{ i'_{k^{'}}j'_{k^{'}}}du, \label{compact}
\end{align}
where $G$ is a compact classical Lie group and $du$ denotes its Haar measure, and $\rho^{\lambda (i)}$, $\bar{\rho}^{\lambda' (i)}$ are irreducible representations and dual representations of $G$ with signatures $\lambda^{(i)}$ and $\lambda^{'(i)}$ respectively. We shall focus on some related duality theorems first and then investigate the asymptotics of this integral as the rank of the group $G$ goes to infinity and the signatures of the representations are fixed. These group integrals as random matrix integrals have important applications in many fields of physics. Let us review some of the physics origins.
\subsection{Group integrals in quantum information theory}
We shall spend more time on quantum information theory than other aspects. In this subsection, let us review some standard definitions and facts that are used in this subsection and the whole paper.  We adopt the following standard conventions
of quantum mechanics: The states of an $n$-level quantum system (and we
include the case infinity) are represented by an n-dimensional complex
Hilbert space $\mathcal{H}$. In this familiar representation, the states of a composite
of two systems A and B say, the first one A $n$-level, and the second B
$m$-level, is then represented by the tensor product of the respective Hilbert
spaces $\mathcal{H}_A$ and $\mathcal{H}_B$. As a result, the composite system (AB) is an (nm)-level
system. This also makes sense in the infinite case where we then use
standard geometry for Hilbert space, and suitable choices of orthonormal
bases (ONBs).
    For a fixed system with Hilbert space $\mathcal{H}$, the corresponding pure quantum
states are vectors in $\mathcal{H}$ of norm one, or rather equivalence classes of such
vectors: Equivalent vectors $v$ and $v'$ in $\mathcal{H}$ yield the same rank-one projection
operator $P$, i.e., the projection of $\mathcal{H}$ onto the one-dimensional subspace in $\mathcal{H}$
spanned by v. In Dirac's terminology, we write $P:= |v \rangle\langle v|$.
 As per Dirac, a bra-ket is an inner product in an ambient Hilbert space $\mathcal{H}$,
while a ket-bra is a rank-one operator in $\mathcal{H}$.
So a bra-ket is a complex scalar, while a ket-bra is an operator.
We will work with the group $\rm U(n)$ of d by d complex unitary matrices; the
matrix entries of a matrix $U$ will be denoted by double subscripts
$U_{ij} = \langle i|U |j \rangle$, and this will refer to the standard ONB in $\mathbb{C}^n$. Normalized
Haar measure will be denoted "$du$". If $n=2$, i.e. $\mathcal{H}\equiv \mathbb{C}^2$, the familiar Pauli matrices offer a conventional realization of pure and mixed states. Set
\begin{align}
\sigma_1=\left(\begin{array}{cc}
        0 & 1 \\
        1 & 0
        \end{array}
        \right),
\sigma_2=\left(\begin{array}{cc}
        0 & -i \\
        i & 0
        \end{array}
        \right),
\sigma_3=\left(\begin{array}{cc}
        1 & 0 \\
        0 & -1
        \end{array}
        \right).
\end{align}
Let $\rho$ be a state, and consider the point
\begin{align}
(\rho(\sigma_1),\rho(\sigma_2),\rho(\sigma_3))\in \mathbb{R}^3.
\end{align}
A computation shows that
\begin{align}
\rho(\sigma_1)^2 +\rho(\sigma_2)^2+\rho(\sigma_3)^2 \leq 1, \label{2}
\end{align}
and the equality holds in (\ref{2}) if and only if the state $\rho$ is pure.
Here the pure state are represented by the points on the two-sphere $S^2\in \mathbb{R}^3$.
Identify a state $\rho$ with
\begin{align}
\rho(A)=\operatorname{Tr}(\rho A),
\end{align}
for all complex $2\times 2$ matrices $A$. The state is pure if and only if $\exists v\in\mathcal{H}$, $\|v\|=1$, such that $\rho=|v\rangle\langle v|$.
   Further, we recall that quantum observables are selfadjoint operators in
H; and states selfadjoint positive semi-definite trace-class operators $\rho$
on H whose trace is one, called density matrices. We can use the
terminology "density matrix" even if $\mathcal{H}$ is infinite-dimensional. By the
Spectral Theorem, a density matrix $\rho$ then corresponds to a pure state if
and only if it is a rank-one projection. In general, a state $\rho$ may be
mixed, in which case it is a convex combination of pure states, allowing for
infinite convex combinations.
    When this standard formalism is applied to the tensor product (AB) of
two quantum systems A and B, then states $\rho$ in the tensor product Hilbert
space have associated marginal states. They are obtained by an application
of a partial trace computation: When we trace over an ONB for the second
system B, i.e., a partial trace-summation applied to $\rho$, we obtain an
associated marginal state $\rho_A$ , where $\rho_A$ is now a density matrix in
the Hilbert space $\mathcal{H}_A$. And analogously, a partial trace summation using an
ONB in $\mathcal{H}_A$ yields $\rho_B$ , the second marginal state.
   Motivated by recent applications, in this paper we are concerned with the
computation of von Neumann entropy of marginal states derived from pure
states in composite systems (AB), formed from tensor factors A and B, A
$n$-level, and B $m$-level.
First, Dirac terminology is defined by
\begin{definition}
$R:=|v\rangle\langle w|$ is a operator that satisfies $Rx:=|v\rangle\langle w | x =\langle w|x\rangle v$, with inner product $\langle w|x\rangle \in \mathbb{C}$.
\end{definition}
Then from the definition, $R^*=|v\rangle\langle w|^* =|w\rangle\langle v|$, and $R^2 =\langle w|v\rangle |v\rangle \langle w|=\langle w|v\rangle R$.
\begin{definition}
Let $\mathcal{H}$ be a complex Hilbert space and mapping the complex Hilbert space $\mathcal{H}_A$ on $\mathcal{H}_B$ such that $\mathcal{H}=\mathcal{H}_A \otimes \mathcal{H}_B$. Let $\rho: \mathcal{H}\rightarrow \mathcal{H}$ be a trace class operator. Pick $\{e_i\}$ ONB in $\mathcal{H}_A$ , and $\{f_j\}$ ONB in $\mathcal{H}_B$. If $x_1,y_1\in \mathcal{H}_A$, set
\begin{align}
\langle x_1| \rho_A y_1\rangle_{\mathcal{H}_A}:=\sum_j \langle x_1\otimes f_j|\rho(y_1\otimes f_j)\rangle_{\mathcal{H}}.
\end{align}
If $x_2,y_2 \in \mathcal{H}_B$, set
\begin{align}
\langle x_2| \rho_B y_2\rangle_{\mathcal{H}_B}:=\sum_i \langle e_i\otimes x_2|\rho(e_i\otimes y_2)\rangle_{\mathcal{H}}.
\end{align}
\end{definition}
\begin{definition}
Let $\mathcal{H}$ be a complex Hilbert space, an set $\mathcal{T}_1(\mathcal{H})=\text{all density matrices }$,i.e., all
$\rho:\mathcal{H}\rightarrow \mathcal{H}$ linear, $\rho$ is a trace class,
$\langle x|\rho x\rangle_{\mathcal{H}}\geq 0,\forall x\in \mathcal{H}$,and $\operatorname{Tr}(\rho)=1$, then by the spectral theorem, the operator $\rho \ln (\rho)$ is well defined. The value $S(\rho):=-\operatorname{Tr}(\rho \ln \rho)$ is called the von Neumann entropy.
\end{definition}
We will compute $S_A(\rho_A)$, and $S_B(\rho_B)$ with $S_{\rho}$. Further note that if $\rho\in \mathcal{T}_1(\mathcal{H})$, there are  projection $P_k:=|v_k\rangle\langle v_k|$, $\lambda_k\in \mathcal{R}, \lambda_k\geq 0, \sum_k \lambda_k=1$, such that
\begin{align}
\rho=\sum_k \lambda_k P_k,
\end{align}
and
\begin{align}
S(\rho)=-\sum_k \lambda_k \ln \lambda_k.
\end{align}
It follows that if $\rho$ is a pure state , then $S(\rho)=0$.
\begin{lemma}
If $\rho\in \mathcal{T}_1(\mathcal{H})$, then the linear functional $\phi:=\phi_\rho$ defined by
\begin{align}
\phi(A):= \operatorname{Tr}(\rho A),
\end{align}
for $A\in B(\mathcal{H})=\text{all bounded operators from $\mathcal{H}$ to $\mathcal{H}$}$,
satisfies
\begin{align}
\phi(A^* A)\geq 0,
\end{align}
and
\begin{align}
\phi(I)=1,
\end{align}
with the identity operator in $\mathcal{H}$.
If $\rho=P_v:=|v\rangle\langle v|$, for $v\in \mathcal{H}$, then
\begin{align}
\operatorname{Tr}(\rho A)=\langle v|A v\rangle_{\mathcal{H}},
\end{align}
for all $A\in B(\mathcal{H})$.
\end{lemma}

\begin{definition}
Let $\rho\in \mathcal{T}_1(\mathcal{H})$. By the spectral theorem, there is an ONB $\{e_i\}$ in $\mathcal{H}$ and an eigenvalue
list $\lambda_1\geq \lambda_2\geq \cdots$, $\sum\lambda_i=1$, such that
\begin{align}
\rho=\sum_{i=1}^\infty \lambda_i |e_i\rangle\langle e_i|,
\end{align}
where $P_i:=|e_i\rangle\langle e_i|$ is the  projection onto $\mathbb{C}e_i$, and $\rho e_i=\lambda_i e_i$.
\end{definition}
If $dim \ (\mathcal{H})=\infty$, then $\lambda_i\rightarrow 0$.
\begin{proof}
An application of the spectral theorem.
\end{proof}

The following lemma is due to Schmit:
\begin{lemma}Let $\forall v\in \mathcal{H}=\mathcal{H}_A\otimes \mathcal{H}_B$, then there exist ONBs $\{e_i\}$ of $\mathcal{H}_A$ and $\{f_i\}$ of $\mathfrak{B}_A$ , and $\xi_k \geq 0$, $\xi_k\geq 0$ such that $v=\sum_k \xi_k e_k\otimes f_k$.
\end{lemma}
\begin{proof}We refer the proof to page 150 in \cite{J2}.
\end{proof}
Remark. The conclusion from the Lemma applies to higher rank tensor products
as well. It follows for example induction: Given an n-fold Hilbert tensor
product H formed from Hilbert spaces $H_i, i = 1, 2, ...n$ as tensor factors;
consider an arbitrary vector $v$ in $\mathcal{H}$ (the tensor product Hilbert space). Then
there are n ONBs, one in each Hilbert space $\mathcal{H}_i$ , the bases depending on the
given vector $v$, and there are numbers $c$ with corresponding indices such that
$v$ has a representation like the case $n=2$, but now with each tensor factor in
the sum being a $n$-fold tensor of vectors from the respective ONBs. And the
convergence with coefficients $c$ holds in the same sense.
\begin{lemma}
Let $\mathcal{H}=\mathcal{H}_A\otimes\mathcal{H}_B$ be a tensor product of Hilbert spaces as described above.
Let $\rho_A\in \mathcal{T}_1(\mathcal{H}_A)$ and $\rho_B\in \mathcal{T}_1(\mathcal{H}_B)$. The the following two conditions are equivalent:\\
(i)The two states $\rho_A$ and $\rho_B$ have the same eigenvalue list.\\
(ii)$\exists \rho\in \mathcal{T}_1(\mathcal{H})$ pure ,such that $\rho_{A}=\operatorname{Tr}_{\mathcal{H}_B} \ (\rho)$ and $\rho_B= \operatorname{Tr}_{\mathcal{H}_A} (\rho)$.
\end{lemma}
\begin{proof}From (ii) to (i).
Suppose a pure state exists $\rho|v\rangle\langle v|\in \mathcal{T}_1(\mathcal{H})$ satisfying the two inequivalent conditions with respect to $\rho_A$ and $\rho_B$. Then $v\in\mathcal{H}$ must satisfy $\|v\|=1$. Since $\mathcal{H}=\mathcal{H}_A\otimes \mathcal{H}_B$, by schmidt's theorem there are ONBs $\{e_i\}$ for $\mathcal{H}_A$ and $\mathcal{H}_B$, and the  there are
\begin{align}
\xi_k\in \mathbb{R}, \ \xi_k\geq 0, \text{ such that} \  v=\sum_k \xi_k e_k\otimes f_k.\label{2}
\end{align}
We will now compute the marginal density matrices $\rho_A$ and $\rho_B$ with the use of the two ONBs in (\ref{2}). From the Schmidt decomposition. Instead if $x,y\in \mathcal{H}_A$, then
\begin{align}
\langle x|\rho_A y\rangle &=\sum_j\langle x\otimes f_j|\rho(y\otimes f_j)\rangle_{\mathcal{H}}=\sum_i \langle x\otimes f_j|v\rangle\langle v|(y\otimes f_j)\rangle_{\mathcal{H}}\nonumber\\
                          &=\sum_{i,k}\xi_i \xi_k \langle x\otimes f_j|e_i\otimes f_i\rangle\langle e_k\otimes f_k| y\otimes f_j\rangle_{\mathcal{H}}\nonumber\\
                          &=\sum_{j,k,k} \xi_i \xi_k \langle x\otimes e_i\rangle_{\mathcal{H}_A}
                          \langle f_j|f_i|f_j\rangle_{\mathcal{H}_B}\langle e_k| y\rangle_{\mathcal{H}_A}\langle f_k|f_j\rangle_{\mathcal{H}_B}\nonumber\\
                          &=\sum_{i,k}\xi_i \xi_k \sum_j \delta_{ji}\delta_{kj}\langle x|e_i\rangle_{\mathcal{H}_A}
                          \langle e_k|y\rangle_{\mathcal{H}_A}=\sum_k \xi_k^2  \langle x|e_k\rangle \langle e_k|y\rangle\nonumber\\
                          &=\langle x|\sum_k \xi_k^2 P_k | y\rangle_{\mathcal{H}_A},
                          \end{align}
with $P_k:=|e_k\rangle \langle e_k |$. It follows that $\{\xi_k^2\}$ is the eigenvalue list for the state
\begin{align}
\rho_A=\sum_k \xi_k^2 P_k. \label{3}
\end{align}
In deed we can arrange the order $\xi_1^2\geq \xi_2^2\geq \cdots $. Since $\|v\|^2_{\mathcal{H}}=1$, it follows
from (\ref{2}) that $\sum_k \xi_k^2=1$. The same argument shows that
\begin{align}
\rho_B=\sum_k \xi_k^2 |f_k\rangle \langle f_k|.  \label{4}
\end{align}
So the same sequence $\{\xi_k^2\}$ form the eigenvalue list of the second marginal state $\rho_B$.

Now let's prove $(i)\Rightarrow (ii)$.

Suppose two states $\rho_A \in \mathcal{T}_1(\mathcal{H}_A)$, $\rho_B\in \mathcal{T}_1(\mathcal{H}_B)$ have the same eigenvalue list $\lambda_k$, $\lambda_1\geq \lambda_2\geq \cdots\geq 0$. Then there are ONBs $\{e_i\}$ in $\mathcal{H}_A$ with
$\rho_A e_i=\lambda_i e_i$; and $\{f_i\}$ in $\mathcal{H}_B$ with $\rho_B f_i=\lambda_i f_i$. Set $v=\sum_i \sqrt{\lambda_i} e_i\otimes f_i$. Then $v\in \mathcal{H}$ satisfies
$\|v\|^2_{\mathcal{H}}=1$; and
$$\operatorname{Tr}_{\mathcal{H}_B} |v\rangle\langle  v|=\rho_A $$, and
$$
\operatorname{Tr}_{\mathcal{H}_A} |v\rangle\langle v|=\rho_B.
$$
Moreover these states satisfy    (\ref{3}) and (\ref{4}) with $|\xi_i|^2 =\lambda_i$.

\end{proof}

 Don N. Page's paper  \cite{P} considered a system $AB$ with Hilbert dimension $mn$. The entropy of a pure state of the whole system is zero. The author  reobtained an approximated formula by of The entropy of the subsystem $A$ which was derived by Lubkin \cite{L}:
\begin{align}
S_{mn}\simeq \ln m-\frac{m}{2n},
\end{align}
and also conjectured
\begin{align}
S_{mn}=\sum_{k=n+1}^{mn}\frac{1}{k}-\frac{m-1}{2n}.
\end{align}
This conjecture was proved later by S.K.Foong and S.Kanno \cite{F}. Page's work is also used in the analysis of the information loss in black hole radiations. It turned out that the information comes out extremely slowly.
Later there have been many work on random pure states of entanglement \cite{G}\cite{S}\cite{A}\cite{B}\cite{L}\cite{C}\cite{G1}\cite{V}\cite{M}.
Therefore it is interesting to compute the average entropy of a random pure state with respect to various symmetries.
The calculation of the entropy will rely on the calculation of  group integrals of unitary group.

\subsection{Group integrals in quantum mechanics and gauge theory}
Weingarten mentioned the asymptotics of the group integrals  are connected to the $g^{-2}$ expansion of the Green's functions of Wilson's formulation of gauge theory on a lattice \cite{W}.  His results showed that how fast the m-string vertices fall in Feynman diagram. Since gauge fields of various representations also involve in the interactions, so the integral (\ref{compact}) is physically important.

Also, in quantum mechanics \cite{S.abers}, when we consider the orbital momentum and spin angular momentum of electrons, we consider the average of the product matrix elements of irreducible representations of $\rm SU(2)$, which is a  more general integral integral:
 \begin{align}
 &\int_{\rm SU(2)}  D^{J_1}_{i_1 j_1}D^{J_2}_{i_2 j_2}\cdots D^{J_k}_{i_k j_k}D^{J^{'}_1 *}_{i^{'}_1 j^{'}_1 }D^{J^{'}_2 *}_{i^{'}_2 j^{'}_2}\cdots D^{J^{'}_{k^{'}} *}_{i^{'}_{k^{'}} j^{'}_{k^{'}}}du ,  \label{su(2)}
\end{align}
where $du$ is the Haar measure of $\rm SU(2)$.
Recall Wigner formula \cite{S.abers}
\begin{align}
D^{J}_{m^{'} m}(\alpha,\beta,\gamma)=e^{-\sqrt{-1}(\alpha m^{'}+\gamma m) }
d^{J}_{m^{'}m}(\beta),
\end{align}
where
\begin{align}
& d^{J}_{m^{'} m}(\beta)\nonumber\\
=& (-1)^{m^{'}-m}\sum_\mu (-1)^\mu \frac{\sqrt{(j+m)!(j-m)!(j+m^{'})!(j-m^{'})!}}{\mu! (j+m-\mu)!(j-\mu-m^{'})!(m^{'}+\mu-m)!}\nonumber\\
&(\cos(\beta/2))^{2j+m-m^{'}-2\mu}(\sin(\beta/2))^{2\mu+m^{'}-m},
\end{align}
and $\mu$ goes over all the possible values that the denominators are defined.
Note that $D^{J}(\alpha,\beta,\gamma)^{*}_{m,m^{'}}=(-1)^{m-m^{'}} D^{J}(\alpha,\beta,\gamma)_{-m,-m^{'}}$.
Apply Wigner formula to (\ref{su(2)}), we get that (\ref{su(2)}) is equal to
\begin{align}
& \frac{1}{32\pi^2}\int_0^{4\pi}d\alpha \int_0^{\pi}\sin\beta d\beta \int_0^{4\pi}\nonumber\\
& e^{-\sqrt{-1}(\alpha(\sum_{p=1}^k i_p-\sum_{p=1}^{k^{'}} i_p^{'})+\gamma(\sum_{p=1}^k j_p-\sum_{p=1}^{k^{'}} j_p^{'})}
(-1)^{\sum_{p=1}^k (i_p-j_p)+\sum_{p=1}^k (i^{'}_p-j^{'}_p)}\nonumber\\
&
(\sum_{\mu,\mu^{'}} (-1)^{\sum_{p=1}^k (\mu_k +\mu^{'}_k) }
\prod_{p=1}^k\frac{\sqrt{(J_k+j_k)!(J_k-j_k)!(J_k+i_k)!(J_k-i_k)!}}{\mu! (J_k+j_k-\mu)!(J_k-\mu-i_k)!(i_k+\mu-j_k)!}\nonumber\\
& \cdot\frac{\sqrt{(J^{'}_k+j^{'}_k)!(J^{'}_k-j^{'}_k)!(J^{'}_k+i^{'}_k)!(J^{'}_k-i^{'}_k)!}}{\mu! (J^{'}_k+j^{'}_k-\mu)!(J^{'}_k-\mu-i^{'}_k)!(i^{'}_k+\mu-j^{'}_k)!}\nonumber\\
& \cdot(\cos(\beta/2))^{\sum_{p=1}^k (2J_k+j_k-i_k-2\mu_k)+\sum_{p=1}^{k^{'}} (2J^{'}_{k^{'}}-j^{'}_{k^{'}}+i^{'}_{k^{'}}-2\mu^{'}_{k^{'}})}\nonumber\\
& \cdot(\sin(\beta/2))^{\sum_{p=1}^k (-j_k+i_k+2\mu_k+\sum_{p=1}^{k^{'}}(j^{'}_{k^{'}}-i^{'}_{k^{'}}+2\mu^{'}_{k^{'}})})\nonumber\\
=& \delta_{0,\sum_{p=1}^k i_p-\sum_{p=1}^{k^{'}} i_p^{'}}
\delta_{0,\sum_{p=1}^k j_p-\sum_{p=1}^{k^{'}} j_p^{'}}\nonumber\\
&
(\sum_{\mu,\mu^{'}} (-1)^{\sum_{p=1}^k (\mu_k +\mu^{'}_k) }
\prod_{p=1}^k\frac{\sqrt{(J_k+j_k)!(J_k-j_k)!(J_k+i_k)!(J_k-i_k)!}}{\mu! (J_k+j_k-\mu)!(J_k-\mu-i_k)!(i_k+\mu-j_k)!}\nonumber\\
& \cdot\frac{\sqrt{(J^{'}_k+j^{'}_k)!(J^{'}_k-j^{'}_k)!(J^{'}_k+i^{'}_k)!(J^{'}_k-i^{'}_k)!}}{\mu! (J^{'}_k+j^{'}_k-\mu)!(J^{'}_k-\mu-i^{'}_k)!(i^{'}_k+\mu-j^{'}_k)!}\nonumber\\
& \int_0^{\pi}  \cos(\beta/2))^{1+\sum_{p=1}^k (2J_k-2\mu_k)+\sum_{p=1}^{k^{'}} (2J^{'}_{k^{'}}-2\mu^{'}_{k^{'}})}
  (\sin(\beta/2))^{1+\sum_{p=1}^k 2\mu_k+\sum_{p=1}^{k^{'}} 2\mu^{'}_{k^{'}}}d\beta)\nonumber\\
=& \delta_{0,\sum_{p=1}^k i_p-\sum_{p=1}^{k^{'}} i_p^{'}}
\delta_{0,\sum_{p=1}^k j_p-\sum_{p=1}^{k^{'}} j_p^{'}}\nonumber\\
&
(\sum_{\mu,\mu^{'}} (-1)^{\sum_{p=1}^k (\mu_k +\mu^{'}_k) }
\prod_{p=1}^k\frac{\sqrt{(J_k+j_k)!(J_k-j_k)!(J_k+i_k)!(J_k-i_k)!}}{\mu! (J_k+j_k-\mu)!(J_k-\mu-i_k)!(i_k+\mu-j_k)!}\nonumber\\
& \cdot\frac{\sqrt{(J^{'}_k+j^{'}_k)!(J^{'}_k-j^{'}_k)!(J^{'}_k+i^{'}_k)!(J^{'}_k-i^{'}_k)!}}{\mu! (J^{'}_k+j^{'}_k-\mu)!(J^{'}_k-\mu-i^{'}_k)!(i^{'}_k+\mu-j^{'}_k)!}\nonumber\\
&  (\sum_{i=0}^{\sum_{p=1}^k J_k-\mu_k+\sum_{p=1}^{k^{'}}J^{'}_{k^{'}}-2\mu^{'}_{k^{'}}}(-1)^i  \frac{2 C^i_{\sum_{p=1}^k J_k-\mu_k+\sum_{p=1}^{k^{'}}J^{'}_{k^{'}}-2\mu^{'}_{k^{'}}}}{i+2+\sum_{p=1}^k 2\mu_k+\sum_{p=1}^{k^{'}}2\mu^{'}_{k^{'}}})).
\end{align}
Then if we let $J_1=J_2=\cdots=J_k=J^{'}_{1}=\cdots=J^{'}_{k^{'}} =J$, and let $J\rightarrow\infty$, then
to understand the asymptotic behavior of (\ref{su(2)}) seems to be a very interesting and difficult problem.

\section{Related duality theorems}
In their paper \cite{CS}, Collins and Sniady offered a beautiful formula to compute the group integral
\begin{align}
\int_{ G} G_{i_1 j_1}G_{i_2 j_2}\cdots G_{i_p j_p} \bar{G}_{i'_1  j'_1}\bar{G}_{i'_2  j'_2 }\cdots \bar{G}_{ i'_p  j'_p }dg, \label{unitary}
\end{align}
where $dg$ is the Haar measure of $G$, and $\bar{G}=(G^{-1})^{T}$. $G$ can be unitary, symplectic or orthogonal group. They reobtained Weingarten's asymptotic formula for this group integral using their formulas. The main tool for their calculation are the following duality theorems.
\begin{theorem}(\cite{Weyl}\cite{GW}\cite{Z})
$V$ is a $N$ dimensional complex vector space. The commutant of the representation $\rho^k(\mathbb{C}[G])\in\operatorname{End}(V^{\otimes k})$ of the group algebra $\mathbb{C}[G]$ on the the tensor space $V^{\otimes k}$ is
\begin{align}
\left\{\begin{array}{cl}
\sigma_k(\mathbb{C}[S_k])    & \text{if} \ G=\rm GL(V),\\
\psi(\mathbb{C}[P_{2k}]\theta_k) & \text{if} \ G=\rm O(V),\\
\psi(\mathbb{C}[P_{2k}]\theta_k) & \text{if} \ G=\rm Sp(V), \text{ only for even $N$},
\end{array}
\right.
\end{align}
where $\sigma_k$ is the natural representation of $S_k$ on $V^{\otimes k}$; $P_{2k}=S_{2k}/\mathfrak{B}_{k}$ is the Brauer algebra with $\mathfrak{B}_{k}=\tilde{S_{k}}\mathcal{R}_k$, $\tilde{S}_k\subset S_{2k}$ is the subgroup of all permutations of the set $\{(1,2),\cdots, (2k-1,2k)\}$, and $\mathcal{R}_k$ is the subgroup generated by the the transformation $2j-1\leftrightarrow 2j$ for $j=1,\cdots, k$; $\theta_k=\psi^{-1}(I_{V^{\otimes k}})$, where $\psi:V^{\otimes 2k}\rightarrow \operatorname{End}(V^{\otimes k})$ is an isomorphism defined by
\begin{align}
\psi(v_1\otimes v_2\otimes\cdots\otimes v_{2k})u=\omega(u,v_2\otimes v_4\otimes\cdots\otimes v_{2k})v_1\otimes v_2\otimes\cdots\otimes v_{2k-1},
\end{align}
where $v_i\in V$, $u\in V^{\otimes k}$, $\omega(u_1\otimes u_2\otimes\cdots\otimes u_k)=\prod_{i=1}^k \omega(u_i,v_i)$. Here $\omega(u_i,v_i)$ is the nondegenerate invariant symmetric bilinear for $\rm O(V)$, or the nondegenerate skew symmetric bilinear form for $\rm Sp(V)$.
\end{theorem}

When we compute the group integral (\ref{compact}), we need to consider the commutant of the group action on the corresponding vector space. Let $W$ be a finite dimensional complex vector space. $\mathcal{A}\subset \operatorname{End}(W)$ is a semisimple algebra,  $\mathcal{A}_1$ is a semisimple subalgebra of $\mathcal{A}$.
By the Double Commutant theorem \cite{GW},
\begin{align}
W\cong\bigoplus_{i}W_{i}\otimes U_i,
\end{align}
where $W_i$ and $U_i$ are irreducible modules of $\mathcal{A}$ and $\mathcal{B}$, i.e.,
$\mathcal{A}=\bigoplus_i \operatorname{End}({W})_i \otimes I_{U_i}$ and $\mathcal{B}=\bigoplus_i I_{{W}_i }\otimes \operatorname{End}(U_i)$.
Restricting the $\mathcal{A}$ representation on $W_{i}$ to $\mathcal{A}_1$, and applying the Double Commutant theorem again, we get
\begin{align}
W\cong\bigoplus_{i,j}W_{ij}\otimes U_{ij}\otimes U_i, \label{dec1}
\end{align}
where $W_{ij}$s are irreducible modules of the representation of $\mathcal{A}_1$ and $U_{ij}$s are the irreducible modules of the commutant of $\mathcal{A}_1$ action on $V_i$. Then
\begin{align}
\mathcal{A}_1=\bigoplus_{i,j}\operatorname{End}(W_{ij})\otimes I_{U_{ij}}\otimes I_{U_i}.
\end{align}
We denote
\begin{align}
\mathcal{B}_{10}=\bigoplus_{i,j} I_{V_{ij}}\otimes \operatorname{End}(U_{ij})\otimes I_{U_i}.
\end{align}
Then (\ref{dec1}) can be written as
\begin{align}
W\cong\bigoplus_{\lambda} W^\lambda\otimes(\bigoplus_{i} U_{i\lambda}\otimes U_i ) , \label{dec2}
 \end{align}
where $\lambda$ runs over all the irreducible representations of $\mathcal{A}_1$ up to equivalent, and $U_{i\lambda}=U_{ij}$ if $W_{ij}\cong W^\lambda$ ($U_{i\lambda}$ is trivial if no such $W_{ij}$).
We choose $0\neq u_{i\lambda}\otimes u_i\in U_{i\lambda}\otimes U_i $ for each nontrivial $U_{i\lambda}\otimes U_i $ and require that $u_{i\lambda}\otimes u_i$ is a basis element of a fixed basis of $U_{i\lambda}\otimes U_i$.
Then define an algebra $\mathcal{B}_{11}=\bigoplus_{\lambda}I_{W^\lambda}\otimes\operatorname{End}(\operatorname{Span}\{u_{i\lambda}\otimes u_i;i\} )$, and mapping any other basis elements in any $U_{i\lambda}\otimes U_i$ to zero. Then the $\mathcal{B}_{11}$ commutes with $\mathcal{A}_1$. Moreover, any nonzero element of $W$ can be mapped to any other elements by the algebra generated by $\{\mathcal{B},\mathcal{B}_{10},\mathcal{B}_{11}\}$.
Therefore we get the following
\begin{lemma}
The commutant of $\mathcal{A}_1$ is generated by $\{\mathcal{B},\mathcal{B}_{10},\mathcal{B}_{11}\}$.
\end{lemma}

Specifically, the commutant of $\rho^k (\mathbb{C}[\rm SU(V)]$ on the vector space $V^{\otimes k}$, where $V$ is a $N$ dimensional complex vector space, is $\sigma_k(\mathbb{C}[S_k])$, which is the same as the commutant of $\rho^k(\mathbb[\rm U(V)]$. This is simply because any matrix $u\in\rm U(V)$ can be written as a complex number multiplying a matrix $g\in \rm SU(V)$.

Now let us consider the commutant of $\rho^k(\mathbb{c}[\rm SO(V)])$ on $V^{\otimes k}$. First of all, recall that if $N$ is even, an irreducible representation  $(\sigma, W)$ of $\rm O(V)$ is determined by a signature $\lambda_1\geq\lambda_2\cdots \lambda_{N/2}\geq 0$:
\begin{align}
\sigma=\left\{\begin{array}{cc}
\rho^\lambda_N & \text{ if $\lambda_{N/2}=0$},\\
\pi^{\lambda,\pm}_N & \text{ if $\lambda_{N/2}\neq 0$}
\end{array}
\right.,
\end{align}
where $\rho^\lambda_N=\operatorname{Ind}_{\rm SO(V)}^{\rm O(V)}$ and $\pi^{\lambda,\pm}_N$ are representations of $\rm SO(V)$ extended to $\rm O(V)$ satisfying $\pi^{\lambda,-}_N= \operatorname{det}\otimes \pi^{\lambda,+}_N$ \cite{GW}.  When $\lambda_{N/2}\neq 0$, the $\rm O(V)$ irreducible representation $\rho^\lambda_N$ decomposes into two $\rm SO(V)$ irreducible representations with signatures $(\lambda_1,\cdots,\lambda_{N/2})$ and $(\lambda_1,\cdots,\lambda_{N/2-1},-\lambda_{N/2})$. The map $\rho(g_0)$ exchanges the two corresponding highest weight and highest weight vectors, where $g_0$ exchanges the two basis elements $e_{N/2}$ and $e_{N/2+1}$ of $V$. By the Double Commutant theorem, the $\rm O(V)$ action on $V^{\otimes k}$ decomposes as
\begin{align}
V^{\otimes k}\cong \bigoplus_{\lambda}\mathcal{R}^\lambda\otimes \Omega^\lambda,\label{dualityo}
\end{align}
where the $\mathcal{R}^\lambda$ is the $\rm O(V)$ irreducible module, and $\Omega^\lambda $ is the $\rm P_{2k}$ irreducible module. We define a $\tilde{\rho}\in\operatorname{End}$ by
\begin{align}
\tilde{\rho}(x)=\left\{\begin{array}{cc}
\rho^k(g_0)(x) & \text{ if $x\in \mathcal{R}^\lambda\otimes\Omega^\lambda$ with $\lambda_{N/2}\neq 0$},\\
x              & \text{ if $x\in \mathcal{R}^\lambda\otimes\Omega^\lambda$ with $\lambda_{N/2}= 0$}.
\end{array}
\right.
\end{align}
To explicitly write the function $\tilde{\rho}(x)$, let us construct the projection operator from $V^{\otimes k}$ to $ \Phi_\lambda$, where $\Phi_\lambda=\mathcal{R}^\lambda\otimes\Omega^\lambda$.
Let $c^{2\lambda}=\sum_{q\in C,p\in R}\operatorname{sgn}(q)qp$ be the Young symmetrizer for the Young diagram with shape $2\lambda_1\geq 2\lambda_2\geq 2\lambda_{[N/2]}\geq 0$, where $C$ is the set of permutations on columns and $R$ is the set of permutations on rows. For a tableau $T$ of this shape, $e_T$ denotes the vector in $V^{\otimes k}$ that corresponds $T$. We shall also use $T$ itself to denote $e_T$. If $e_T=v_1\otimes v_2\otimes\cdots\otimes v_{2k}$, we require that in the tableau $T$, all the odd indexes are in the odd columns and $v_{2i}$ is always right after $v_{2i-1}$, $1\leq i\leq k$.
\begin{proposition}
$\epsilon^\lambda=\sum_{s\in S_{2k}}\frac{1}{\mu^2}\psi(s c^{2\lambda} s^{-1} \theta_k)$ is the projection from $V^{\otimes k}$ to $\mathcal{R}^\lambda\otimes\Omega^\lambda$.
\end{proposition}
\begin{proof}
Each space $\epsilon^\lambda (V^{\otimes k})$ is an invariant subspace of $\psi(P_{2k}\theta_k)$. For different signatures $\lambda^{(1)}$ and $\lambda^{(2)}$, $\epsilon^{\lambda^{(1)}}\epsilon^{\lambda^{(2)}}=0$, by the property of Young symmetrizer. The number of these subspaces is equal to the number of the $\mathcal{R}^\lambda\otimes\Omega^\lambda$.
Moreover, note that the generating function for $\rm SO(V)$ irreducible representation with signature $\lambda$ is $\alpha(g)=\prod_{i=1}^{[N/2]}\Delta^{r_i}$, where $\Delta_i$s are principal minors of $g\in\rm SO(V)$,
$r_i=\lambda_i-\lambda_{i+1}$, for $1\leq i\leq [N/2]-1$, and $r_{[N/2]}=\lambda_{[N/2]}$. Therefore
$c^\lambda(T^\lambda_0)$, where tableau $T^\lambda_0$ of shape $\lambda_1\geq \lambda_2\geq\cdots\geq\lambda_{[N/2]}\geq 0$ with all 1s in the first row, all 2s in the second row, $\cdots$, $[N/2]$ in the $[N/2]$th row, is a highest weight vector of $\rm O(V)$(other indexes are $\bar{1}$, $\bar{2}$, $\cdots$ satisfying $\omega(i,\bar{j})=\delta_{i j}$ and $\bar{\bar{i}}=i$. For $\rm O(V)$, if $N$ is odd, the index of $e_N$ is 0 with $\bar{0}=0$). Recall that for any $\sigma\in S_{2k}$, $\phi(\sigma\theta_k)$, it is the product of two operators The first which is the product of some operators of the form $D_{ij}C_{ij}$, for different pairs $(i,j)$, $1\leq i<j\leq k$ (the trivial case is identity operator) and the second is an operator which is an element of the representation of $S_k$ on $V^{\otimes k}$ \cite{GW}. Here $C_{ij}:V^{\otimes k}\rightarrow V^{\otimes k-2}$ is called $ij$-contraction operator defined by
\begin{align}
C_{ij}(v_1\otimes v_2\otimes\cdots\otimes v_k)=
\omega(v_i,v_j)v_1\otimes\cdots\otimes\hat{v}_i\otimes\cdots\hat{v}_j\otimes\cdots \otimes v_k,
\end{align}
where $v_i$ and $v_j$ are omitted,
and $D_{ij}:V^{\otimes k-2}\rightarrow V^{\otimes k}$ is defined by
\begin{align}
D_{ij}(v_1\otimes v_2\otimes\cdots\otimes v_k)=\sum_{p=1}^N v_1\otimes v_2\otimes\cdots f_p\otimes\cdots\otimes f^p\otimes\cdots\otimes v_{k-2},
\end{align}
where $f_p$ and $f^p$ are at the $i$th and $j$th positions respectively and $\{f_p\}$ is a basis and $\{f^p\}$ is its dual basis via $\omega$.
Note than any $\omega(e_i,e_j)=0$ for $i,j=1,2,\cdots,[N/2]$.  Hence in the expression of $\psi(c^{2\lambda}\theta_k)(T_0)$, all the row permutations containing nontrivial normalized Brauer diagram vanishes (chapter 10, \cite{GW}). Therefore $\psi(c^{2\lambda}\theta_k)(T_0)$ is a constant multiplying $c^\lambda(T_0)$. Then $\epsilon^\lambda (V^{\otimes k})$ contains the vector space of highest weight vectors of irreducible representation of $\rm O(V)$ with signature $\lambda_1\geq\lambda_2\geq\cdots\geq\lambda_{[N/2]}\geq 0$. Therefore $\epsilon^\lambda$ is the projection operator.
\end{proof}
We would like to remark that this proposition and the proof apply to symplectic groups too.

Then we have proved the following
\begin{theorem}
If $N$ is odd, the commutant of $\rho^k(\mathbb{C}[\rm SO(V)])$ is $\rho^k(\mathbb{C}[SO(V)])$; if $N$ is even, the commutant is  $\operatorname{Span}\{\psi(\mathbb{C}[P_{2k}]\theta_k),\psi(\mathbb{C}[P_{2k}]\theta_k)\tilde{\rho}\}$.
\end{theorem}

In general, denote the action of the group algebra $\mathbb{C}[G]$ on a vector space $W$ simply by $\rho^k(\mathbb{C}[G])$. If $\bar{W}$ is the dual vector space of $W$, the commutant $\operatorname{End}_G W$ of $\rho^k(\mathbb{C}[G])$ can be identified as the subspace of invariant tensors in the tensor space $W\otimes \bar{W}$ \cite{GW}, which is the vector space which is the direct sum of one dimensional irreducible subspaces in the irreducible decomposition of $W\otimes \bar{W}$ under the group $G$. The method of $Z$ invariants will give all the irreducible subspaces \cite{Z}.

\section{The asymptotic behaviors of classical group integrals}
In his paper \cite{W}, Weingarten obtained the following asymptotics:\\
For $\rm U(N)$,
\begin{align}
& \int_{\rm U(N)} U_{i_1 j_1}U_{i_2 j_2}\cdots U_{i_q j_q} U^{*}_{i^{'}_1 j^{'}_1}
U^{*}_{i^{'}_2 j^{'}_2}
\cdots U^{*}_{i^{'}_q j^{'}_q}du \nonumber\\
&=\frac{1}{N^q}
\sum_{\sigma\in S_q}\delta_{i_1 i^{'}_{\sigma(1)}}\delta_{j_1 j^{'}_{\sigma(1)}}\cdots
\delta_{i_q i^{'}_{\sigma(q)}}\delta_{j_q j^{'}_{\sigma(q)}}+O(\frac{1}{N^{q+1}}). \label{1}
\end{align}
For $\rm SO(N)$,
\begin{align}
& \int_{\rm SO(N)}U_{i_1 j_1}U_{i_2 j_2}\cdots U_{i_{2q} j_{2q}}du \nonumber\\
& =\frac{1}{N^q}\sum \delta_{i_{k_1} i_{l_1}}\cdots \delta_{i_{k_q} i_{l_q}} +O(\frac{1}{N^{q+1}}), \label{2}
\end{align}
where the sum carries over all the partitions of $\{1,2,\cdots, 2q\}$ into pairs $(k_1,l_1),\cdots,(k_q,l_q)$.
For $\rm Sp(2N)$,
\begin{align}
& \int_{\rm Sp(2N)}U^{k_1}_{i_1 j_1}U^{k_2}_{i_2 j_2}\cdots U^{k_{2q}}_{i_{2q} j_{2q}}du \nonumber\\
& =\frac{1}{(2N)^q}\sum
M^{k_{l_1} k_{m_1}}_{i_{l_1} i_{m_1}}M^{k_{l_1} k_{m_1}}_{j_{l_1} j_{m_1}}\cdots M^{k_{l_q} k_{m_q}}_{i_{l_q} i_{m_q}}M^{k_{l_q} k_{m_q}}_{j_{l_q} j_{m_q}} +O(\frac{1}{(2N)^{q+1}},) \label{3}
\end{align}
 where $U^1_{ij}=U_{ij}$,$U^2_{ij}=U^*_{ij}$, $M^{kl}_{ij}=J_{ij}$ if $k=l$, $M^{kl}_{ij}=\delta_{ij}$ if $k\neq l$,
$J_{ij}=-\delta_{i,(j-1)}$ if $i$ is even, and $J_{ij}=\delta{i,(j+1)}$ if $i$ is odd.

In their paper\cite{CS} Benoit Collins,Piotr Sniady offered a method to compute the integral (\ref{unitary}), i.e., for the vector representation of unitary group $\rm U(N)$, $\rm SO(N)$, and $\rm SpU(N)$. By the virtue of our duality theorems, this method also theoretically applies to the integrals of irreducible representations of group $G$. Here we offer an more transparent and elementary method to compute the group integral (\ref{unitary}) for unitary, orthogonal and symplectic groups.
Denote the following integral
\begin{align}
\int_G \rho^\lambda_{i_1 j_1}\cdots \rho^\lambda_{i_q j_q} \bar{\rho}^\lambda_{i^{'}_1 j^{'}_1}\cdots \bar{\rho}^\lambda_{i^{'}_q j^{'}_q}du,
\end{align}
by Dirac notation
\begin{align}
P_G(I,J,I^{'},J^{'})=\int_{G} \langle I |\rho^\lambda (u) |J\rangle\langle J^{'}|\rho^{\lambda }(u^{-1})|I^{'}\rangle du, \label{Dirac}
\end{align}
where $\rho^\lambda$ denotes the irreducible representation of $G$ with signature $\lambda$, $|J\rangle=|e_{j_1}\otimes \cdots e_{j_q}\rangle$ with $J=(j_1,\cdots, j_q)$.
Then $|J\rangle\langle I^{'}|\in \operatorname{End}(\mathbb{C}^{N_\lambda})^{\otimes p}$, where $N_\lambda$ is the dimension of the irreducible representation $\rho^\lambda$ of $G$.
we can define
\begin{definition}
\begin{align}
\Phi^{\lambda}_G(J,J^{'}):=\sum_{\sigma\in B }\operatorname{Tr}(|J\rangle\langle J^{'}|\sigma^{-1})\sigma,
\end{align}
where $B$ is the finite group that generates the group algebra $\mathcal{B}$, which is the commutant of the group algebra $\rho^{\otimes q}(\mathbb{C}[G])$, and $\sigma$ also denotes the representation.
\end{definition}
The existence  of $B$ is a straightforward consequence of the Double Commutant theorem.
The conditional expectation is
\begin{align}
\mathbb{E}^{\lambda}_G(J,J^{'})=\int_{G} \rho^{\lambda}(u) |J\rangle\langle J^{'}| \rho^{\lambda}(u^{-1}) du,
\end{align}
where $du$ is the Haar measure of $G$, $\rho^{\lambda}_G(u)\in\operatorname{End}(\mathbb{C}^{N_\lambda})^{\otimes q}$ is the action of $u\in G$ on the tensor space $(\mathbb{C}^{N_\lambda})^{\otimes q}$,  $|J\rangle=|e_{j_1}\otimes e_{j_2}\otimes\cdots e_{j_q}|$ and $\langle J^{'}|=\langle e_{j^{'}_1}\otimes e_{j^{'}_2}\otimes\cdots e_{j^{'}_q}|$.
Then we can compute $P_G(I,I^{'},J,J^{'})$ by
\begin{proposition}
\begin{align}
\Phi^\lambda_G(J,J^{'})=\mathbb{E}^\lambda_G(J,J^{'})\Phi^\lambda_G(\operatorname{Id}),
\end{align}
$\Phi^\lambda_G(\operatorname{Id})$ has an inverse.
\end{proposition}
\begin{proof}
Since $\operatorname{Tr}\circ \mathbb{E}^\lambda_G=\operatorname{Tr}$, it can be easily checked that $\Phi^\lambda_G$ is a $\mathcal{B}$ bimodule and
\begin{align}
\Phi^\lambda_G(J,J^{'})=\mathbb{E}^\lambda_G(J,J^{'})\Phi^\lambda_G(\operatorname{Id}).
\end{align}
Taking $J=J^{'}=e_1\otimes e_2 \otimes\cdots\otimes e_q$, we get
\begin{align}
\operatorname{Id}=\mathbb{E}^\lambda_G(|e_1\otimes e_2 \otimes\cdots\otimes e_q\rangle\langle e_1\otimes e_2 \otimes\cdots\otimes e_q|)\Phi^\lambda_G(\operatorname{Id}).
\end{align}
\end{proof}
However, in order to get the asymptotics of (\ref{compact}) we need the following decomposition lemma.
Let us fix the signatures and let $n\rightarrow \infty$, because $\operatorname{Par}(N,k)=\operatorname{Par}(N,k)$, when $p\geq N$. We consider $\lambda^{(i)}=(m^{(i)}_1,m^{(i)}_2,\cdots,m^{(i)}_{M^{(i)}},0,0,\cdots)$ of an irreducible representation, where $m^{(i)}_1\geq m^{(i)}_2,\cdots\geq m^{(i)}_{M^{(i)}}\geq 0$ are integers, for $G=\rm U(N)$, $\rm SO(N)$, or $\rm SpU(2N)$ when $N$ is large. Let us consider the asymptotic behavior of the integral (\ref{compact}) with respect to $N$.
Let us assume $G=\rm U(N)$ first. Recall that an irreducible module of $\rm U(N)$(also $\rm GL(N,\mathbb{C})$ with signature $\lambda=(m_1,m_2,\cdots,m_{M})$) can be constructed by $\rm GL$ standard tableaux of shape $\lambda$ with elements in $\{1,2,\cdots,N\}$ \cite{Fulton}. $\rm GL(N)$ standard tableaux are the tableaux which rows are nondecreasing and which columns are increasing.
The Young symmetrizer $c$ is defined by  $c=\sum \pm qp$ (see \cite{Z}), where $q$ ranges over the column permutation $C$, and $p$ ranges over the row permutations $R$ of the Young diagram. The sign is $+$ or $-$ according to whether $q$ is even or odd. It is well known that the $\{c(T);  T \text{ is $\rm GL(N)$ standard Tableaux of shape $\lambda$}\}$ gives a basis of the irreducible representation of $\rm U(N)$ with signature $\lambda$.
Then  we can construct an ONB of the irreducible $\rm U(N)$ module. For other classical groups, there are also corresponding standard tableaux. Then we get the following decomposition(not irreducible).
\begin{lemma}$G$ is a classical group. There is a decomposition under the group action of $G$:
\begin{align}
 & \bigotimes_{i=1}^k \mathbb{C}^{N_{\lambda^{(i)}}}\otimes  \bigotimes_{i=1}^{k'}\mathbb{C}^{N_{\lambda^{'(i)}}}\nonumber\\
& \simeq
 \bigotimes_{i=1}^k \operatorname{Span}\{ c(T); T \text{ is $G$ standard Tableaux of shape $\lambda^{(i)}$}\}\nonumber\\
& \otimes \bigotimes_{i=1}^{k^{'}} \operatorname{Span}\{ c(\bar{T}); T\text{ is $G$ standard Tableaux of shape $\lambda^{' (i)}$}\},
\end{align}
where $N_\lambda$ is the dimension of the irreducible representation $\rho^\lambda_G$ of $G$; $\bar{T}$ is the tableau replacing every index in $T$ by its dual index.
\end{lemma}

Let us come back to $G=\rm U(N)$. For a signature $\lambda$, denote $\tilde{e}_T=\frac{1}{n_T}c(T)$, for $\forall T\in\mathbb{T}^\lambda$, where $n_T $ is the normalization constant such that $\|\tilde{e}_T \|=1$($\|\cdot\|$ is defined by the inner product $\langle \cdot, \cdot\rangle$  of the Hilbert space $(\mathbb{C}^{N_\lambda})^{\otimes k}$). Then $\{\tilde{e}_T\}$ with $T$ runs over all $\rm GL(N)$ standard tableaux is an ONB of the irreducible $\rm U(N)$ module. Then $ n_T^2=\langle c(T),c(T)\rangle$. Since $d=\frac{1}{\mu}c$ is a projection (see chapter 8,\cite{Z}). Then $n_T^2=\mu \langle c(T)|T\rangle$. Note that the $\mu$ and multiplicity $t$ of the representation $\rho^\lambda_{\rm U(N)}$ in $\Phi_m\simeq (\mathbb{C}^n)^{\otimes m}$  satisfy the relation $\mu\cdot t=m!$, where $m=\sum_{i=1}^n m_i$ (Theorem 1 in Chapter 8 \cite{Z}). On the other hand, by the RSK  correspondence, $t$ is the number of the tableaux of shape $\lambda$ with entries from 1 to $N$ each occurring once \cite{F}. A consequence of hook formula shows that $t=\frac{m!\prod_{i<j}(l_i-l_j)}{l_1! \cdot l_2!\cdot\cdots \cdot l_n!}$, where $m=\sum_{i=1}^n m_N$, $l_i=m_i+N-i$, $i=1,2,\cdots,N$. For each term of $c(T)$, unless it is $T$ itself, its inner product with $T$ will be zero. Moreover, the only transformations of the form $pq$, where $p\in R$, and $q\in C$ that fixe $T$ are those $q=1$ and $p$ makes each row unchanged. Let $m_{\mu\nu}$, $1\leq\mu\leq\nu\leq N$, denotes the Gelfand diagram that corresponds the tableau $T$(if $(\mu,\nu)$ is not in this range, then $m_{\mu\nu}$ is set to be zero). Then the number of these transformations is
\begin{align}
f=\prod_{\mu,\nu}(m_{\mu\nu}-m_{\mu,\nu-1})!
\end{align}

We use $T(i)$ to denote the map from an ONB element of an irreducible $G$ module indexed by $i$ to the corresponding $G$ standard tableau.

Let $W$ be any vector space. We generalize the Kronecker delta function to be a bilinear form on $W^{\otimes k}\times W^{\otimes k}$.
\begin{definition}
The bilinear form $\delta_{\cdot,\cdot}$ on $W^{\otimes k}\times W^{\otimes k}$ is defined by
\begin{align}
\delta_{e_{i_1}\otimes e_{i_2}\cdots e_{i_k} ,e_{i^{'}_1}\otimes e_{i^{'}_2}\cdots e_{i^{'}_k}}
=\delta_{i_1 i^{'}_1}\delta_{i_2 i^{'}_2}\cdots\delta_{i_k i^{'}_k}
\end{align}
\end{definition}

 Therefore we have the following
\begin{theorem}
If $G=\rm U(N)$, the group integral (\ref{compact}) with signatures $\lambda^{(i)}=(m^{(i)}_1,m^{(i)}_2,\cdots, m^{(i)}_n)$ with $m^{(i)}_1\geq m^{(i)}_2\geq \cdots\geq m^{(i)}_n\geq 0$ is equal to
\begin{align}
&\prod_{p=1}^k \frac{1}{n_{T(i_p)}n_{T(j_p)}}\prod_{p=1}^{k^{'}}\frac{1}{n_{T(i^{'}_p)}n_{T(j^{'}_p)}}
(\sum_{p_{1,i},p_{2,i},p^{'}_{1,i},p^{'}_{2,i} \in R, q_{1,i}, q_{2,i},q^{'}_{1,i}, q^{'}_{2,i} \in Q}  \prod_{i=1}^k \operatorname{sgn}(q_{1,i}) \operatorname{sgn}(q_{2,i})\cdot\nonumber\\
&\int_{\rm U(N)} du \langle q_{1,1} p_{1,1} T(i_1)|\rho^{\otimes m^{(1)}}_{\rm U(N)} q_{2,1}p_{2,1}T(j_1)\rangle \cdots\langle q_{1,k}p_{1,k}T(i_k)|\rho^{ \otimes m^{(k)}}_{\rm U(N)}q_{2,k}p_{2,k} T(j_k)\rangle\nonumber\\
& \langle q_{1,1} p_{1,1} T(i^{'}_1)|\rho^{* \otimes m^{(1)}}_{\rm U(N)} q_{2,1}p_{2,1}T(j^{'}_1)\rangle
\cdots\langle q^{'}_{1,k^{'}}p^{'}_{1,k^{'}}T(i^{'}_{k^{'}})|\rho^{* \otimes m^{(k^{'})}}_{\rm U(N)} q^{'}_{2,k^{'}}p^{'}_{2,k^{'}}T(j^{'}_{k^{'}})\rangle).
\end{align}
Its asymptotic behavior with respect to $N$ is
\begin{align}
& \prod_{p=1}^k \frac{1}{n_{T(i_p)}n_{T(j_p)}}\prod_{p=1}^{k^{'}}\frac{1}{n_{T(i^{'}_p)}n_{T(j^{'}_p)}}(\frac{1}{N^m}\sum_{\sigma\in S_m,p_{1,i},p_{2,i},p^{'}_{1,i},p^{'}_{2,i} \in R, q_{1,i}, q_{2,i},q^{'}_{1,i}, q^{'}_{2,i} \in Q}   \prod_{i=1}^k \operatorname{sgn}(q_{i})\operatorname{sgn}q^{'}_{i^{'}}\nonumber\\
& \delta_{q_{1,1}p_{1,1}T(i_1)\otimes \cdots \otimes q_{1,k}p_{1,k}T(i_k), \sigma q_{1,1}p_{1,1}T(i^{'}_1)\otimes\cdots \otimes q_{1,k}p_{1,k}T(i_1)}\nonumber\\
& \delta_{q_{2,1} p_{2,1 } T(j_{1})\otimes\cdots\otimes q_{2,k} p_{2,k} T(j_{k}),\sigma q_{2,1} p_{2,1 } T(j^{'}_{1})\otimes\cdots\otimes q_{2,k} p_{2,k } T(j^{'}_{k^{'}})}+O(\frac{1}{N^{m+1}})),
\end{align}
where $m^{(r)}=\sum_l m^{(r)}_l$,$m=\sum_{r=1}^k m^{(r)}$. $m^{'}$ is similarly defined. $m=m^{'}$, otherwise the integral vanishes.
\end{theorem}

If $G=\rm O(N)$, we need to construct a ONB for the irreducible $\rm O(N)$ modules. Proctor \cite{Proctor}, King and Welsh \cite{KW} constructed irreducible $\rm O(N)$ modules. They defined $\rm O(N)$ standard tableaux satisfying the following conditions. Let $w_{\bar{i}}=e_{2i-1}$, $w_{i}=2i$, for $i=1,2,\cdots,[N/2]$. If $N$ is odd, $w_0=e_{N}$. The orthogonal standard tableaux defined in \cite{Proctor} is
\begin{definition}(\cite{Proctor})
Let $\lambda$ be a partition of $N$ such that $\tilde{\lambda}_1+\tilde{\lambda}_2\leq N$ and $T^\lambda_{ab}$ denotes the entry of the $a$th row and $b$th column of the tableau. For $i=1,2,\cdots, r$ with $r=[N/2]$, let $\alpha_i$ and $\beta_i$ be the numbers of the entries less than or equal to $i$ in the first and second columns, respectively of the tableau $T^\lambda$. $T^\lambda$ is $\rm O(N)$ standard tableau if and only if it is $\rm GL(N)$ standard tableau and
for each $i=1,2,\cdots,r$,\\
(i) $\alpha_i+\beta_i\leq 2i$; \\
(ii)If $\alpha_i +\beta_i=2i$ with $\alpha_i>\beta_i$ and $T^\lambda_{\alpha_i,1}=\bar{i}$ and $T^\lambda_{\beta_i b}=i$ for some $b$ then $T^\lambda_{\beta_i-1,b}=\bar{i}$;\\
(iii)If $\alpha_i +\beta_i=2i$ with $\alpha_i=\beta_i=i$ and $T^\lambda_{i,1}=\bar{i}$ and $T^\lambda_{\beta_i b}=i$ for some $b$ then $T^\lambda_{i-1,b}=\bar{i}$.
\end{definition}
If $T$ is a $\rm O(N)$ standard tableau, $T_{0}$ denotes the quotient of the Young symmetrized tableau $\{T\}$ with the subpace of the form
\begin{align}
\sum_{i\in\mathcal{I}} x\otimes w_i\otimes y\otimes w_{\bar{i}}\otimes z,
\end{align}
where $x,y,z$ are arbitrary tensor powers of  $V$.
Similar to  Weyl's theorem \cite{Weyl} for the other definition of $\rm O(N):=\{M;MM^{T}=I, M\in\rm GL(N)\}$, the tensor space $V^{\otimes m}$ can be decomposed into the direct sum
of traceless subspace and its complement subspace which is spanned by all the tensors of the above form and these two subspaces are orthogonal to each other, i.e.,
$V^{\otimes m}=V^{\otimes m}_0+V^{\otimes m}_{1}$, where
$V^{\otimes m}_0$ is the subspace such that for any two indexes $v_i,v_j$ of $v=\cdots\otimes x\otimes\cdots\otimes y\otimes \cdots\in V^{\otimes m}_0$,
\begin{align}
\sum_{i\in\mathcal{I}} \langle w_{\bar{i}}|v_i\rangle \langle v_j|w_i\rangle x\otimes\hat{v}_i\otimes y\otimes\hat{v}_j\otimes z
=0,\label{traceless}
\end{align}
and
\begin{align}
& V^{\otimes m}_{1}\nonumber\\
=&\operatorname{Span}\{\sum_{i\in\mathcal{I}} x\otimes w_i\otimes y\otimes w_{\bar{i}}\otimes z; x, y, z \text{ are tensor powers of }V\} \label{trace}
\end{align}
is its complement. Therefore $T_{0}\in V^{\otimes m}_0$. Use $T_{1}$ to denote $T-T_{0}$. $T_{0}$ can be computed by taking traces for all the pairs of indexes of the equation (\ref{traceless}).
If $T$ is a $\rm O(N)$ standard tableau, then it is straightforward to check $ (cT)_0=cT_{0}$ and $(cT)_{1}=c T_{1}$, where $c$ is the Young symmetrizer.  Let $T$ and $T^{'}$ be two $\rm O(N)$ standard tableaux. We shall compute $\langle cT_0|cT^{'}_0\rangle$. 

If $T=T^{'}$, $\langle cT_0|cT^{'}_0\rangle=\mu\langle T|cT_0\rangle=\mu \langle cT- (cT)_{1}|T \rangle$. Therefore it suffices to compute $\langle (cT)_{1}|T \rangle$. By computing the traces of $T$, and solving a linear system, we get that
\begin{align}
T_{1}=\left\{\begin{array}{lc}
\frac{1}{N-1+\sum_{\mu=1}^{N/2}  (s_{\mu}-s_{\mu-1})( s_{\bar{\mu}}-s_{\bar{\mu}-1})}\sum_{1\leq i<j \leq m }D_{ij}C_{ij}T &  \text{if $N$ is even},\\
\frac{1}{N-1+C^2_{s_n-s_{N-1}} +\sum_{\mu=1}^{(N-1)/2}  (s_{\mu}-s_{\mu-1})( s_{\bar{\mu}}-s_{\bar{\mu}-1})}\sum_{1\leq i<j \leq m}D_{ij}C_{ij}T &  \text{if $N$ is odd},
\end{array}
\right. \label{T1}
\end{align}
where $s_{\nu}=\sum_{\mu=1}^{\nu} m_{\mu\nu}$, $C_{ij}$ and $D_{ij}$ are $ij$-contraction and $ij$-expansion operators. One can check that for any $p\in R$, $q\in C$, only those such that $qp T=T$ or $qpT$ is switching two entries of $T$ can make $\langle (qpT)_{1}|T\rangle$ nonvanishing. The only $qp$ such that $qpT=T$ are those with $q=I$, so the number of these transformations is $\prod_{\mu,\nu}(m_{\mu\nu}-m_{\mu,\nu-1})!$. The number of the transformations of switching two entries consists of two types of transformations: the first type is that switching two entries $\nu$ and $\bar{\nu}$ in the same column, which is equal to
\begin{align}
\sum_{\mu} f\prod_{\nu< \bar{\nu}}
(m_{(\mu+1)\bar{\nu}}-m_{\mu (\nu-1)})_{+},
\end{align}
where $x_{+}=\left\{\begin{array}{cc}
x & \text{if $x>0$,}\\
0 & \text{otherwise}.
\end{array}
\right.$
The second type is that switching two entries $\nu$ and $\bar{\nu}$ in the same row. The second type also consists of two kinds of transformations: one is those that have identity $q$, and the other one is those that have nonidentity $q$. The number of the first kind is
\begin{align}\sum_{\mu} f\prod_{\nu_1< \bar{\nu}_1<
\nu_2< \bar{\nu}_2} (m_{\mu \nu_1}-m_{\mu, (\nu_1-1)})(m_{\mu \nu_2}-m_{\mu, (\nu_2-1)}).
\end{align}
The number of the second kind is
\begin{align}
& \sum_{\mu} f\prod_{\nu< \bar{\nu}} (m_{(\mu+1)\bar{\nu}}-m_{\mu (\nu-1)})_{+}(m_{\bar{\mu}\nu}-m_{\mu\nu})\nonumber\\
& +\sum_{\mu} f\prod_{\nu< \bar{\nu}} (m_{\mu\bar{\nu}}-m_{(\mu-1),(\nu-1)})_{+} (m_{\mu\nu}-m_{\mu,(\nu-1)}),
\end{align}
if the $\operatorname{sign}(q)=-1$; and
\begin{align}
\sum_{\mu} f\prod_{\nu< \bar{\nu}}
(m_{(\mu+1)\bar{\nu}}-m_{\mu (\nu-1)})_{+}(m_{\mu \bar{\nu}}-m_{(\mu-1),(\nu-1)})_{+},
\end{align}
if the $\operatorname{sign}(q)=1$.
Then for the sum of $\langle \operatorname{sign}(q)(qpT)_1, T\rangle$ for all $qpT$ switching two indexes $\nu$ and $\bar{\nu}$.
If $qpT=T$, $\langle (qpT)_1|T\rangle$ can be explicitly computed using (\ref{T1}).
Then $\langle cT_0|cT_0\rangle$ can be computed explicitly by plugging in all the above five formulas.

If $T\neq T^{'}$, and the sets of the entries of $T$ and $T'$ are the same, then there exists some $s\in S_k$, such that $T=sT'$, and $s\neq qp$ for any $q\in C$, $p\in R$. By the property of Young symmetrizer, $\langle cT_0|cT^{'}_0\rangle=\langle cT_0|c_s T_0\rangle=\mu\langle T_0|c c_sT_0\rangle=0$. If the sets of the entries of $T$ and $T'$ are the same except one entries, then it is easy to see that $\mu\langle T|cT^{'}_0\rangle=0$. If the sets of the entries of $T$ and $T'$ are the same except two entries, if
$\langle T|T^{'}_1\rangle\neq 0$, then the two entries in $T$ and $T'$ must be $i$,$\bar{i}$ and $j$, $\bar{j}$ respectively and $i\neq j,i\neq\bar{j}$ and we are able to compute $\langle T|cT^{'}_1\rangle=\langle T|\sum \operatorname{sgn}(q)qp T^{'}_1\rangle=\frac{1}{N}|\{p; pT^{'}=T^{'}\}|$, where $|\{p; pT^{'}=T^{'}\}|$ is the cardinality of the set $\{p; pT^{'}=T^{'}\}$(this is easy to compute depending on the positions of the two entries). Then we are able to compute $\langle cT_0|cT^{'}_0\rangle$. If more than two elements in the sets of entries of $T$ and $T^{'}$ are different, $\langle cT_0|cT^{'}_0\rangle$ is always zero. After we compute all the inner products among $cT_0$, where $T$ runs over $\rm O(N)$ standard tableaux, we are able to compute the ONB of the irreducible $\rm O(N)$ module. The basis elements of the ONB are denoted $\tilde{T}(i)$, $1\leq i\leq N_\lambda$, where $N_\lambda$ is the dimension of the irreducible $\rm O(N)$ module. Now we have the following
\begin{theorem}
If $G=\rm O(N)$, the group integral
$$
 \int_{\rm O(N)}  \rho^{\lambda^{(1)}}_{i_1 j_1}\rho^{\lambda^{(2)}}_{ i_2 j_2}\cdots \rho^{\lambda^{(k)}}_{ i_k j_k}du
$$
with signatures $\lambda^{(i)}=(m^{(i)}_1,m^{(i)}_2,\cdots, m^{(i)}_n)$ with $m^{(i)}_1\geq m^{(i)}_2\geq \cdots\geq m^{(i)}_n\geq 0$, $i=1,\cdots,k$, is equal to
\begin{align}
&\int_{\rm O(N)} \langle \tilde{T}(i_1)|\rho^{ m^{(1)}}_{\rm O(N)}\tilde{T}(j_1)\rangle \cdots\langle \tilde{T}(i_k)|\rho^{ m^{(k)}}_{\rm O(N)}\tilde{T}(j_k)\rangle du.
\end{align}
Its asymptotic behavior with respect to $N$ is
\begin{align}
& \frac{1}{N^{m/2}}\sum_{\sigma\in P_{m}} 
\delta_{ \tilde{T}(i_1)\otimes\cdots\otimes\tilde{T}(i_k),\sigma \tilde{T}(i_1)\otimes\cdots\otimes \tilde{T}(i_1)}
\delta_{\tilde{T}(j_{1})\otimes\cdots\otimes\tilde{T}(j_{k}),\sigma   \tilde{T}(j_{1})\otimes\cdots\otimes  q_{2,k} \tilde{T}(j_{k})}+O(\frac{1}{N^{m/2+1}}),
\end{align}
where $m^{(r)}=\sum_l m^{(r)}_l$, and $m=\sum_{r=1}^k m^{(r)}$ is even. If $m$ is odd, the integral vanishes.
\end{theorem}
For $G=\rm SO(N)$, the formula can be derived using the $\rm SO(N)$ standard tableau
in \cite{KW}.

If $G=\rm SpU(2N)$, similar to the orthogonal group,  the entries of tableaux are ordered as $\bar{1}<1< \bar{2}<2<\cdots <\bar{N}<N$. The $\rm Sp(2N)$ standard tableau is defined such that if the entries are increasing in each column., non-decreasing in each row and if, in addition, the elements of row $i$ are all greater than or equal to $\bar{i}$, for each $i$ \cite{KE}\cite{Berele}. The ONB $\tilde{T}_{i}$, $1\leq i\leq N_\lambda$, where $N_\lambda$ is the dimension of the irreducible $\rm Sp(2N)$ module, can be computed following similar steps as the orthogonal group case, noting that the tensor space $V^{\otimes m}$  decomposes into different subspaces:
$V^{\otimes m}=V^{\otimes m}_{\tilde{0}}+V^{\otimes m}_{\tilde{1}}$, where
$V^{\otimes m}_{\tilde{0}} $
 is the traceless subspace such that for any two indexes $v_1,v_2$ of $v=\cdots\otimes x\otimes\cdots\otimes y\otimes \cdots\in V^{\otimes m}_{\tilde{0}}$,
$$\sum_{i=1}^n (\cdots\otimes \langle w_{\bar{i}}|v_1\rangle\otimes\cdots\otimes \langle v_2|w_i\rangle \otimes \cdots-\cdots\otimes \langle w_{{i}}|v_1\rangle\otimes\cdots\otimes \langle v_2|w_{\bar{i}}\rangle \otimes \cdots
=0,
$$
and
\begin{align}
& V^{\otimes m}_{\tilde{1}}\nonumber\\
=& \operatorname{Span}\{\sum_{i\in\mathcal{I}} x\otimes w_i\otimes y\otimes w_{\bar{i}}\otimes z-x\otimes w_{\bar{i}}\otimes y\otimes w_i\otimes z; x, y, z \text{ are tensor powers of }V\}\nonumber \label{traceless}
\end{align} is its complement.

Let $W$ be any vector space. We define a bilinear form on $W^{\otimes k}\times W^{\otimes k}$:
\begin{definition}
The bilinear form $M_{\cdot,\cdot}$ on $W^{\otimes k}\times W^{\otimes k}$ is defined by
\begin{align}
M(e_{i_1}\otimes e_{i_2}\otimes\cdots\otimes e_{i_k},e_{i^{'}_1}\otimes e_{i^{'}_2}\otimes\cdots\otimes e_{i^{'}_k})
=\prod_{p=1}^k M(e_{i_p},e_{i^{'}_p}),\nonumber
\end{align}
where
\begin{align}
M_{e_{i_p},e_{i^{'}_p}}=\left\{\begin{array}{cc}
J_{i_p,i^{'}_p}   &   \text{if $i_p$ and $i^{'}_p$ are both covariant or contravariant}, \\
\delta_{i_p,i^{'}_p } & \text{otherwise}.
\end{array}
\right.\nonumber
\end{align}
\end{definition}
Then we can state the following
\begin{theorem}
If $G=\rm SpU(N)$, the group integral (\ref{compact}) with signatures $\lambda^{(i)}=(m^{(i)}_1,m^{(i)}_2,\cdots, m^{(i)}_n)$ with $m^{(i)}_1\geq m^{(i)}_2\geq \cdots\geq m^{(i)}_n\geq 0$ is equal to
\begin{align}
&\int_{\rm SpU(N)} du \langle \tilde{T}(i_1)|\rho^{m^{(1)}}_{\rm SpU(2N)} \tilde{T}(j_1)\rangle\cdots\langle \tilde{T}(i_k)|\rho^{m^{(k)}}_{\rm SpU(2N)}\tilde{T}(j_k)\rangle)\nonumber\\
& \langle  \tilde{T}(i^{'}_1)|\rho^{*  m^{(1)}}_{\rm SpU(2N)}\tilde{T}(j^{'}_1)\rangle
\cdots\langle \tilde{T}(i^{'}_{k^{'}})|\rho^{*  m^{(k^{'})}}_{\rm SpU(2N)}\tilde{T}(j^{'}_{k^{'}})\rangle.
\end{align}
Its asymptotic behavior with respect to $N$ is
\begin{align}
& \frac{1}{(2N)^{m/2}}\sum_{\sigma\in P_{m}}
M_{\tilde{T}(i_1)\otimes\cdots\otimes \tilde{T}(i^{'}_{k^{'}}), \sigma \tilde{T}(i_1)\times\cdots\otimes \tilde{T}(i^{'}_{k^{'}})}
M_{\tilde{T}(j_1)\otimes\cdots\otimes \tilde{T}(j^{'}_{k^{'}}), \sigma \tilde{T}(j_1)\times\cdots\otimes \tilde{T}(j^{'}_{k^{'}})}+\nonumber\\
& O(\frac{1}{(2N)^{m/2+1}}),
\end{align}
where $m^{(r)}=\sum_{l} m^{(r)}_l$, and $m=\sum_{r=1}^k m^{(r)}+\sum_{r=1}^{k^{'}} m^{'(r)}$ is even. If $m$ is odd, the integral vanishes when $n>m$.
\end{theorem}
Then we complete the computations of the asymptotic behaviors of (\ref{compact}) for fixed signatures.

% ----------------------------------------------------------------

\end{document}